\documentclass[preprint,aps,prd,showpacs]{revtex4}

\usepackage{tabularx}
\usepackage{dcolumn}
\usepackage{graphics}
\usepackage{bm}
\usepackage[dvips]{graphicx}
\usepackage{tabularx}
\usepackage{amsmath}
\usepackage{amssymb}
\usepackage{longtable}
\usepackage{verbatim}
\usepackage{float}
\usepackage{fancyhdr}

\begin{document}

\title{The Effect of the Earth Matter on Three Neutrino Oscillations and
Sensitivity to CP Phase Parameter}

\author{Bushra Shafaq}
\email{bushra.chep@pu.edu.pk}
\thanks{}
\author{Faisal Akram}
\email{faisal.chep@pu.edu.pk}
\thanks{}

\affiliation{Centre For High Energy Physics, Punjab University,
Lahore(54590), Pakistan.}
\date{\today }
\begin{center}
\date{\today }
\end{center}

\begin{abstract}
We find an analytical expression of neutrino evolution operator in
the Earth matter using perturbative approach in the context of three
neutrino oscillations. We find that our analytical expression is
highly accurate by comparing its results with the numerical
solutions of neutrino evolution equation at energy scales relevant
for solar, reactor, atmospheric, and accelerator neutrinos. Using
our analytical approach we study the accuracy of hypothesis of
treating the Earth density piecewise constant. We also study how the
Earth matter effect can change the sensitivity to CP phase parameter
$\delta _{CP}$. Through nadir angle averaged conversion
probabilities of neutrino and anti-neutrino, we find that the
sensitivity to $\delta _{CP}$ is maximum in the energy range 0.2 to
1 GeV and through energy averaged conversion probabilities, we find
that the sensitivity is maximum about nadir angle $72.5^{\text{o}}$
for neutrinos oscillating in the Earth matter.
\end{abstract}

\keywords{Neutrino oscillations, Leptonic CP phase, Solar and
reactor neutrinos, Atmospheric and accelerator neutrinos}

\pacs{14.60.Pq, 26.65.+t, 13.15.+g, 91.35.-x}

\maketitle
\section{Introduction}

Solar neutrino problem (SNP) came into sight in 1968 when Homestake chlorine
experiment \cite{homestake} found that the measured flux of solar neutrinos $%
v_{e}$ was significantly smaller than predicted by the standard solar model
(SSM). An elegant way to explain the observed depletion, already proposed by
Pontecorvo \cite{pontecarvo}, was the neutrino oscillations in which a
neutrino of one flavor transforms into another while propagating in free
space. This lead to an extensive experimental program in neutrino physics to
confirm the depletion and hypothesis of neutrino oscillations. The
subsequent Ga experiments (SAGE \cite{sage}, GALLEX \cite{gallex}, and GNO
\cite{GNO}), which like Homestake also measured the flux of solar neutrinos $%
v_{e}$ through charge current (CC) weak interaction, confirmed the
depletion. Kamiokande \cite{kamiokande} and later Super-Kamiokande (SK) \cite%
{SK,SK1,SK2,SK3,SK4} experiments pioneered in real-time solar neutrinos
observation and provided the direct evidence that solar neutrinos are coming
from the direction of sun, a feature which greatly helped in separating the
signal from background. Both Kamiokande and SK measured solar neutrinos
through elastic scattering process $v_{x}+e\rightarrow v_{x}+e$, which is
sensitive to all flavors of active neutrinos $(x=e,\mu ,$ and $\tau )$,
however sensitivity to $\nu _{\mu }$ and $\nu _{\tau }$ is reduced because $%
\sigma (v_{\mu ,\tau }e)\approx 0.16\sigma (v_{e}e)$. Consequently the
detectors were unable to measure total flux of all flavors of neutrinos
coming from the direction of sun, a quantity which was crucial to confirm
the hypothesis of neutrino oscillations. The direct confirmation of solar
neutrino hypothesis was provided by a new real-time solar neutrino
experiment SNO (Sudbury Neutrino Observatory) \cite{SNO}, when it published
the result \cite{SNO1} of $^{8}$B solar neutrinos flux measured by neutral
current (NC)\ weak interaction process $(v_{x}+d\rightarrow v_{x}+p+n)$.
This NC\ process is equally sensitive to all neutrino flavors, hence the
measured flux is sum of the flux of all neutrino flavors. The agreement with
SSM prediction provided a conclusive prove of neutrino oscillations in solar
neutrinos.\ The next step was to determine neutrino oscillations parameters.
Neutrino oscillations can occur if it is assumed that flavor states (the
states which take part in CC and NC weak interactions)\ are different from
mass eigen states and at least one neutrino is massive. The unitary matrix
which relates flavor and mass eigen states is defined through 4 parameters,
in which 3 are the angles $(\theta _{12},\theta _{23},\theta _{13})$ and one
is the CP phase parameter $\delta _{CP}$ \cite{3nu}, assuming neutrino are
Dirac particles. The survival or conversion probabilities of neutrinos
additionally depend on two independent squared mass differences $(\Delta
m_{21}^{2},\Delta m_{31}^{2})$. So there are total six constants which
describe neutrino oscillations. In case where only two flavor of neutrinos
are assumed, the number the independent parameters are reduced to two; one
mixing angle and one squared mass difference. To calculate solar neutrino $%
\nu _{e}$ data, we merely require the knowledge of averaged electron
neutrino survival probability, which in three neutrino oscillations depends
on the mixing angles $\theta _{12}$ and $\theta _{13}$, and squared mass
difference $\Delta m_{21}^{2}$. The global analysis of solar neutrino data
of SK+SNO experiments \cite{ga1} showed, LMA (large mixing angle) solution
with $\Delta m_{21}^{2}=4.8\times 10^{-5}$ eV$^{2},\sin ^{2}\theta
_{12}=0.31 $ is the best solution. However, other solutions were not ruled
out with sufficient statistical significance. In the Ref. \cite{ga1} the
constraint $\sin ^{2}\theta _{13}=0.0219$ coming from reactor neutrino
experiments is assumed, which shows that dependence of survival probability
on $\theta _{13} $ is extremely weak so that effectively 2 neutrino
oscillations are sufficient to describe solar neutrino data. The issue of
finding unique solution of mixing parameters was resolved by KamLAND
experiment of reactor anti-neutrinos $\overline{\nu }_{e}$, which is
sensitive to $\Delta m^{2}\sim 10^{-5}$. The global analysis of solar
neutrino data of SK+SNO and KamLAND\ experiment \cite{kland1} finally
identified the LMA solution with $\Delta m_{21}^{2}=7.49\times 10^{-5}$ eV$%
^{2},\sin ^{2}\theta _{12}=0.307$ as a true solution of solar neutrino
problem. Neutrino oscillations are also studied in atmospheric neutrino
experiments. Atmospheric neutrinos predominantly $v_{\mu }$ and $\overline{v}%
_{\mu }$ are produced by the decay of $\pi $ and $K$ mesons, which are
provided by primary cosmic rays in upper atmosphere. Neutrino oscillations
in the atmospheric neutrinos were first discovered by observing muon
neutrinos disappearance by SK \cite{SKa}. Later experiments, MACRO \cite%
{macro}, Soudan 2 \cite{soudan 2}, and MINOS \cite{minos} confirmed the
observation. Unfortunately their exist large uncertainties in the
predictions of atmospheric neutrinos flux, which makes it difficult to
precisely determine the neutrino oscillation parameters. More recently the
muon disappearance is also observed in long-baseline accelerator neutrino $%
\nu _{\mu }$ experiments, K2K \cite{k2k}, MINOS+ \cite{minosp}, T2K \cite%
{t2k}, and NO$\nu $A \cite{nova}. Electron neutrino appearance $\nu _{\mu
}\rightarrow \nu _{e}$ is measured by MINOS \cite{minose}, whereas
anti-electron appearance $\overline{\nu }_{\mu }\rightarrow \overline{\nu }%
_{e}$ is meaured by T2K \cite{t2ke}. These measurements allow precise
determination of the parameters $\theta _{23}$ and $\Delta m_{32}^{2}$.
There exist several different analysis, here we report the fitted values
given in PDG \cite{PDG}; $\sin ^{2}\theta _{23}=0.417_{-0.028}^{+0.025}$, $%
\Delta m_{32}^{2}=2.51\pm 0.05(10^{-3}$ eV$^{2})$, assuming normal mass
hierarchy (i.e., $m_{1}<m_{2}<m_{3}$). The parameter $\sin ^{2}\theta _{13}$
is extracted from the measured disappearance of anti-neutrinos $\overline{%
\nu }_{e}$ in reactor neutrinos experiments (Double Chooz \cite{doublechooz}%
, RENO \cite{reno}, and Daya Bay \cite{dayabay}), at relatively small
distance $L\sim 1$ km\thinspace\ corresponding to large value of $\Delta
m_{32}^{2}$. Fitted average value reported in PDG is $\sin ^{2}\theta
_{13}=2.12\pm 0.08(10^{-2})$. Having relatively precise knowledge of
neutrino oscillations parameters, the interest is now shifted to CP-phase $%
\delta _{CP}$. Recently $\delta _{CP}=1.45_{-0.26}^{+0.27}\pi $ is measured
by T2K experiment \cite{T2Kcp} through the study of difference between
conversion probabilities of $v_{\mu }\rightarrow v_{e}$ and $\overline{v}%
_{\mu }\rightarrow \overline{v}_{e}$ using accelerator neutrinos.

Although neutrinos are weakly interacting particles, their propagation
through matter significantly modifies the neutrino oscillations through
coherent interaction with matter electrons via charge changing (CC) weak
interaction \cite{matcon}. In case of matter of constant density, the
parameters of oscillations are effectively changed, depending upon the value
of energy and density. Dependence of effective oscillating parameters on
energy and density has a resonance character \cite{matcon}, which can lead
to a strong enhancement of the oscillations, independent of the values of
vacuum mixing angles. However, when conversion probability is averaged over
energy, the maximum depletion which it can yield is 1/2. A large observed
depletion (less than 1/2) in solar neutrinos $\nu _{e}$ flux is the
consequence of MSW\ (Mikheyev--Smirnov--Wolfenstein) effect \cite{msw} which
occurs in a slowly varying density, while crossing the point where resonance
condition is satisfied in the interior of Sun. MSW effect, produced by the
Earth's variable density, is not relevant for solar, atmospheric,
accelerator, and reactor neutrinos, as for reactor and solar neutrinos
resonance density become very large and for atmospheric and accelerator
neutrinos it becomes very small as compared to electron density in the
Earth. Nevertheless it is found that the Earth's density profile can
significantly affect the neutrino oscillations. For atmospheric or
accelerator neutrinos, having energies $E\gtrsim 2$ GeV, the effect of
relatively small value of $\Delta m_{21}^{2}=7.49\times 10^{-5}$ eV$^{2}$
can be neglected to leading order, consequently the Earth matter strongly
suppresses the oscillations due to $\Delta m_{21}^{2}$. In this case three
neutrino oscillations probabilities are effectively described by two
neutrino oscillations. The relevant oscillation parameters are $\theta _{13}$%
, $\Delta m_{31}$, and Earth electron density $N_{e}$. It is shown in Refs.
\cite{penha,penha1,penha2}, the probability $P(v_{e}\rightarrow v_{\mu (\tau
)})$ is maximally enhanced for neutrino of energy $E\simeq 7.1$ GeV and
traveling the path length $L\simeq 11740$ km for $\Delta m_{31}^{2}\simeq
2.5\times 10^{-3}$ eV$^{2}$, $\sin ^{2}2\theta _{13}\simeq 0.09$, and $%
N_{e}=2.2N_{A}$ cm$^{-3}$. The result is established by using the analytical
solution of neutrino evolution equation through the Earth. This parametric
enhancement due to the effect of the Earth matter can significantly amplify
oscillation probabilities for both atmospheric and accelerator neutrinos as
the parametric values are close to LMA solution. It is also noted that this
resonance like effect, though amplifying the conversion $v_{e}\rightarrow
v_{\mu (\tau )}$, suppresses $\overline{v}_{e}\rightarrow \overline{v}_{\mu
(\tau )}$ if $\Delta m_{31}^{2}>0$. For $\Delta m_{31}^{2}<0$, the effect is
reversed and amplification is produced for $\overline{v}_{e}\rightarrow
\overline{v}_{\mu (\tau )}$. Solar neutrinos can also be affected by the
coherent interaction with the Earth matter. Solar neutrinos detected at
night time reach the detector after passing through the Earth interior,
which produces a small enhancement in $\nu _{e}$ flux through $v_{\mu (\tau
)}\rightarrow v_{e}$. This regeneration effect for solar neutrinos has been
observed in SK, SNO, and BOREXINO \cite{BOREX,BOREX1} through non-zero value
of day-night asymmetry $A_{D-N}$ of measured event rate.

The effect of parametric enhancement is studied using analytical expression
of neutrino oscillations obtained by treating the Earth density piecewise
constant. Usually divided into two regions; core $(0\leq r<3485$ km$)$ and
mantle $(3486\leq r<6371$ km$)$. In this work we obtain an analytical
expression of neutrino evolution operator treating the Earth density
piecewise variable in five shells. For each shell the variation is treated
perturbatively about its average value. This scheme is also adopted in Ref.
\cite{2v}, where the problem is solved for two neutrino oscillations. Our
analytical expressions agree with numerical solutions of three neutrino
evolution equation in the Earth at energies relevant for solar, reactor,
atmospheric, and accelerator neutrinos. We also study, how the Earth matter
effect can change the sensitivity to CP phase parameter $\delta _{CP}$

In Sec. II, we describe the general formalism of 3 neutrino
oscillations. In Sec. III, we discuss the solution of evolution
equation in matter of constant density. In Sec. IV, we discuss the
parametrization of radial profile of electron density in the Earth.
In Sec. V, we apply perturbation theory to obtain solution of
evolution operator. In Sec. VI, we discuss accuracy of our
analytical expressions and study the effect of the Earth matter on
neutrino oscillations and sensitivity to CP phase. \

\section{The general Formulism}

Neutrino flavor states $|\nu _{\alpha }\rangle $ are written as linear
combinations of mass eigen states $|\nu _{k}\rangle $ \cite{pontecarvo} via
a unitary matrix, called Pontecorvo-Maki-Nakagawa-Sakata (PMNS) matrix, as
following
\begin{equation}
{|\nu _{\alpha }\rangle =\sum_{k=1}^{3}U_{\alpha k}^{\ast }|\nu _{k}\rangle ,%
}
\end{equation}%
where $\alpha =e,\mu ,\tau $ and $k=1,2,3$. We use the following
parametrization of PMNS matrix \cite{parmeter}
\begin{equation}
{U=\left(
\begin{array}{ccc}
c_{12}c_{13} & s_{12}c_{13} & s_{13}e^{i\delta } \\
-s_{12}c_{23}-c_{12}s_{23}s_{13}e^{-i\delta } &
c_{12}c_{23}-s_{13}s_{13}s_{23}e^{-i\delta } & c_{13}s_{23} \\
s_{12}s_{23}-c_{12}s_{13}c_{23}e^{-i\delta } &
-c_{12}s_{23}-s_{12}s_{13}c_{23}e^{-i\delta } & c_{13}c_{23}%
\end{array}%
\right) ,}
\end{equation}%
where $c_{ij}=\cos \theta _{ij}$, $s_{ij}=\sin \theta _{ij}$ for $i=1,2,3$
and $\delta _{CP}$ is the Dirac CP phase. For anti-neutrinos, CP phase is
replaced by $-\delta _{CP}$. An arbitrary neutrino state $|\psi (t)\rangle $
can be expressed in terms of both flavor or mass eigen states.
\begin{equation}
{|\psi (t)\rangle =\sum_{\alpha }\psi _{\alpha }(t)|\nu _{\alpha }\rangle
=\sum_{k}\psi _{k}(t)|\nu _{k}\rangle ,}
\end{equation}%
where $\psi _{\alpha }(t)$ and $\psi _{k}(t)$ are the components of
the state $|\psi (t)\rangle $ in flavor and mass basis respectively
and they are related as
\begin{equation}
{\psi _{k}(t)=\sum_{\alpha }U_{\alpha k}^{\ast }\psi _{\alpha }(t).}
\label{5}
\end{equation}%
In matrix form Eq. \ref{5} is written as following
\begin{equation}
{\psi ^{f}(t)=U\psi ^{m}(t),}
\end{equation}%
where
\begin{equation}
{\psi ^{f}(t)=\left(
\begin{array}{c}
\psi _{e}(t) \\
\psi _{\mu }(t) \\
\psi _{\tau }(t)%
\end{array}%
\right) ,}\text{ \ \ \ \ \ \ }{\psi ^{m}(t)=\left(
\begin{array}{c}
\psi _{1}(t) \\
\psi _{2}(t) \\
\psi _{3}(t)%
\end{array}%
\right) .}
\end{equation}%
In mass basis, the vacuum Hamiltonian is
\begin{equation}
{H_{0}^{(m)}=\left(
\begin{array}{ccc}
E_{1} & 0 & 0 \\
0 & E_{2} & 0 \\
0 & 0 & E_{3}%
\end{array}%
\right) .}
\end{equation}%
Whereas in flavor basis it is given by ${H_{0}^{(f)}=UH_{0}^{(m)}U^{\dag }}$%
. As neutrinos are relativistic, having very small masses as compared to
their energies, so the energy $E_{i}\simeq p+\frac{m_{i}^{2}}{2p}$.
Subtracting the constant $p+\frac{m_{1}^{2}}{2p}$ from the diagonal of ${%
H_{0}^{(m)}}$%
\begin{equation}
{H_{0}^{(m)}=}\frac{1}{2p}{\left(
\begin{array}{ccc}
0 & 0 & 0 \\
0 & \Delta m_{21}^{2} & 0 \\
0 & 0 & \Delta m_{31}^{2}%
\end{array}%
\right) ,}
\end{equation}%
\noindent where $\Delta m_{ij}^{2}=m_{i}^{2}-m_{j}^{2}$ and $E\simeq p$.
When neutrinos are propagating through matter, the total Hamiltonian in
flavor basis is sum of vacuum Hamiltonian $H_{0}^{(f)}$ and an interacting
part $H_{I}^{(f)}(x)$.%
\begin{equation}
{H^{(f)}(x)=H_{0}^{(f)}+H_{I}^{(f)}(x).}
\end{equation}%
Flavor states of neutrinos ($\nu _{e}$, $\nu _{\mu }$, and $\nu _{\tau })$
interact with the electrons in the matter through charge current (CC) and
neutral current (NC) weak interactions. Since only $\nu _{e}$ neutrinos take
part in CC interaction and all flavors are equally sensitive to NC
interaction, therefore NC interaction amplitudes do not contribute to $%
H_{I}^{(f)}(x)$. It is noted that decoherent interaction, in which the state
of incoming neutrino is changed, has negligible effect on the propagation of
solar, atmospheric, accelerator, and reactor neutrinos. The interaction
Hamiltonian is, therefore, given as
\begin{equation}
{H_{I}^{(f)}(x)=A(x)\left(
\begin{array}{ccc}
1 & 0 & 0 \\
0 & 0 & 0 \\
0 & 0 & 0%
\end{array}%
\right) ,}
\end{equation}%
where $A(x)=2\sqrt{2}G_{F}N_{e}(x)$, in which $G_{F}$ is Fermi coupling
constant and $N_{e}(x)$ is electron density in matter. For anti-neutrinos
the sign of electron density is inverted in the interaction Hamiltonian.%
\newline
For calculational convenience we also convert $H^{(f)}(x)$ into a traceless
matrix $\tilde{H}^{(f)}(x)$ defined as following
\begin{equation}
\tilde{H}^{(f)}{(x)=H^{(f)}(x)-\frac{1}{3}}\mathrm{Tr}{(H^{(f)}(x))I,}
\label{hft}
\end{equation}%
where
\begin{equation}
\mathrm{Tr}{(H^{(f)}(x))=}\frac{1}{2p}\left( \Delta m_{21}^{2}+\Delta
m_{31}^{2}\right) {+A(x).}
\end{equation}

\section{\noindent Solution of time evolution operator in a constant density
matter}

\noindent The time evolution equation of neutrino in matter is given as
\begin{equation}
{\frac{i\partial \psi ^{f}(t)}{\partial t}=\tilde{H}^{(f)}(x)\psi ^{f}(t).}
\label{evo}
\end{equation}%
In matter of constant density, $\tilde{H}^{(f)}$ does not depends upon $x$
so its solution is given as
\begin{equation}
{\psi ^{f}(t)=e^{-i\tilde{H}^{(f)}t}\psi ^{f}(0).}
\end{equation}%
Neutrinos travel with speed very close to the speed of light so $t=x$. Thus
the time evolution operator is given by
\begin{equation}
{\mathcal{U}(x)=e^{-i\tilde{H}^{(f)}x}.}
\end{equation}%
In order to obtain a computationally useful expression of $\mathcal{U}(x)$,
we follow the Ref. \cite{tommy}, in which $\tilde{H}^{(f)}$ is decomposed
into linear combination of Gell-Mann matrices $(\lambda _{j})$ as following
\begin{equation}
{\ \tilde{H}^{(f)}=h_{j}\lambda _{j},}
\end{equation}%
where $h_{j}=\frac{1}{2}\mathrm{Tr}(\tilde{H}^{(f)}\lambda _{j})$ are real
coefficients obtained by applying orthonormalization condition $Tr[{\lambda
_{i}\lambda _{j}]=2\delta }_{ij}$. Similarly time evolution operator can
also be expresses as a linear combination of Gell-Mann matrices and identity
matrix, which is required because $\mathcal{U}(x)$ is not traceless.
\begin{equation}
{\mathcal{U}(x)=u_{0}I+iu_{j}\lambda _{j},}  \label{ut}
\end{equation}%
where $u_{0}=\frac{1}{3}\mathrm{Tr}[\mathcal{U}(x)]$ and $u_{j}=\frac{1}{2i}%
\mathrm{Tr}[\mathcal{U}(x){\lambda _{j}}]$ are again obtained by using the
orthonormalization condition of Gell-Man matrices. These coefficients can be
expressed in terms of eigen values of ${\tilde{H}^{(f)}}$, which we
represent by $E_{\alpha }^{(m)}$, as following.
\begin{subequations}
\begin{eqnarray}
u_{0} &=&\frac{1}{3}\sum\limits_{\alpha =1}^{3}e^{-iE_{\alpha }^{(m)}x},
\label{u0} \\
u_{j} &=&-\frac{i}{2}\sum\limits_{a=1}^{3}\frac{\partial E_{\alpha }^{(m)}}{%
\partial h_{j}}e^{-iE_{\alpha }^{(m)}x},  \label{uj}
\end{eqnarray}%
\noindent where the second equation is obtained by using $\frac{\partial {%
\mathcal{U}}}{\partial h_{j}}=-ix\mathcal{U}{\lambda _{j}}$. The eigen
values of the matrix $\tilde{H}^{(f)}$ are given by following characteristic
polynomial equation.
\end{subequations}
\begin{equation}
E_{\alpha }^{(m)3}+a_{2}E_{\alpha }^{(m)2}+a_{1}E_{\alpha }^{(m)}+a_{0}=0
\end{equation}%
where $a_{2}=-\mathrm{Tr}[\tilde{H}^{(f)}]=0$, $a_{1}=-\frac{1}{2}(\mathrm{Tr%
}[\tilde{H}^{(f)}]^{2}-\mathrm{Tr}[(\tilde{H}^{(f)})^{2}])=h_{i}h_{i}\equiv
\left\vert h\right\vert ^{2}$, and $a_{0}=-\det [\tilde{H}^{(f)}]=-\frac{2}{3%
}d_{jkl}h_{j}h_{k}h_{l}$, in which totally symmetric tensor $d_{jkl}=\frac{1%
}{4}\mathrm{Tr}[\{{\lambda _{j},\lambda _{k}\}}\lambda _{l}]$.
Differentiating the characteristic equation with respect to $h_{j}$ to yield
$\frac{\partial E_{\alpha }^{(m)}}{\partial h_{j}}$, given as%
\begin{equation}
{\frac{\partial E_{\alpha }^{(m)}}{\partial h_{j}}=\frac{2\left( E_{\alpha
}^{(m)}h_{j}+[h\ast h]_{j}\right) }{3(E_{\alpha }^{(m)})^{2}-\left\vert
h\right\vert ^{2}},}
\end{equation}%
\noindent where $[h\ast h]_{j}=d_{jkl}h_{k}h_{l}$. Using it in Eq. (\ref{uj}%
) to obtain $u_{j}\,$and then ${\mathcal{U}(x)}$ from Eq. (\ref{ut})
\begin{equation}
\mathcal{U}(x)=\frac{1}{3}\sum_{\alpha =1}^{3}e^{-iE_{\alpha
}^{(m)}t}A_{\alpha },  \label{fut1}
\end{equation}%
where $A_{\sigma }=(I+\rho _{\alpha }\chi _{\alpha j}\lambda _{j})$,
$\rho _{\alpha }=\frac{2}{3E_{\alpha }^{(m)}-|h|^{2}}$ and $\chi
_{\alpha j}=E_{\alpha }^{(m)}h_{j}+[h\ast h]_{j}$. For completeness
we also give the solution of characteristic equation
\begin{subequations}
\small
\label{eigen}
\begin{align}
E_{1}^{(m)}& =-\sqrt{\frac{a_{1}}{3}}\cos \left[ \frac{1}{3}\tan ^{-1}\left(
\frac{1}{a_{0}}\sqrt{-a_{0}^{2}+\frac{4a_{1}^{3}}{27}}\right) \right] +\sqrt{%
a_{1}}\sin \left[ \frac{1}{3}\tan ^{-1}\left( -\frac{1}{a_{0}}\sqrt{%
a_{0}^{2}+\frac{4a_{1}^{3}}{27}}\right) \right] , \\
E_{2}^{(m)}& =-\sqrt{\frac{a_{1}}{3}}\cos \left[ \frac{1}{3}\tan ^{-1}\left(
\frac{1}{a_{0}}\sqrt{-a_{0}^{2}+\frac{4a_{1}^{3}}{27}}\right) \right] -\sqrt{%
a_{1}}\sin \left[ \frac{1}{3}\tan ^{-1}\left( -\frac{1}{a_{0}}\sqrt{%
a_{0}^{2}+\frac{4a_{1}^{3}}{27}}\right) \right] , \\
E_{3}^{(m)}& =2\sqrt{\frac{a_{1}}{3}}\cos \left[ \frac{1}{3}\tan ^{-1}\left(
\frac{1}{a_{0}}\sqrt{-a_{0}^{2}+\frac{4a_{1}^{3}}{27}}\right) \right] .
\end{align}%
\normalsize
\end{subequations}
Notice that $\sum_{\alpha =1}^{3}E_{\alpha }^{(m)}=0$, which follows
from the fact that $\tilde{H}^{(f)}$ is traceless. The eigen values
given by Eqs.
(\ref{eigen}) can be used in Eq. (\ref{fut1}) along with $h_{j}=\frac{1}{2}%
\mathrm{Tr}(\tilde{H}^{(f)}\lambda _{j})$ to determine the evolution
operator ${\mathcal{U}(x)}$ for the case of constant density.

\section{Parameterizing the Earth electron density}

\noindent Electron density in the Earth is provided by Preliminary
Earth Reference Model (PREM) \cite{prem}, which divides the Earth
interior into 8 shells of continuous density. We follow the scheme
of Ref. \cite{2v} in which 4 outer shells are grouped into single
shell and the density in each shell is fitted using following
polynomial
\begin{equation}
{N(r)=\alpha _{k}+\beta _{k}r^{2}+\gamma _{k}r^{4},}  \label{nr}
\end{equation}%
where $k=1$ to $5$, are the labels of the shells. The fitted values
of the coefficients are given in Ref. \cite{2v} and we summarized
them in Table 1. The functional form of Eq. (\ref{nr}) is invariant
for non radial neutrino trajectory (i.e., nadir angle $\eta \neq 0$)
\begin{equation}
{N_{k}(x)=\alpha _{k}^{\prime }+\beta _{k}^{\prime }x^{2}+\gamma
_{k}^{\prime }x^{4},}
\end{equation}%
where
\begin{subequations}
\begin{align}
{\alpha _{k}^{\prime }}& {=\alpha _{k}+\beta _{k}\sin ^{2}\eta +\gamma
_{k}\sin ^{4}\eta ,} \\
{\beta _{k}^{\prime }}& {=\beta _{k}+2\gamma _{k}\sin ^{2}\eta ,} \\
{\gamma _{k}^{\prime }}& {=\gamma _{k}.}
\end{align}%
The trajectory coordinate $x$ of neutrino and nadir angle $\eta $ are
defined in the Fig. 1. In each shell the density is split as following
\end{subequations}
\begin{equation}
{N_{k}(x)=\bar{N_{k}}+\delta N_{k}(x),}  \label{nx}
\end{equation}%
where $\bar{N_{k}}$ is the ($\eta $ dependent) average density along
the shell chord and $\delta N_{k}(x)$ is the residual density, which
can be obtained by using equation (\ref{nx}) itself for given
$\bar{N_{k}}$ obtained as following
\begin{equation}
{\bar{N_{k}}=\int_{x_{k}-1}^{x_{k}}dxN_{k}(x)/(x_{k}-x_{k}-1).}
\end{equation}%
\begin{table}[h]
\begin{center}
\begin{tabular}{|c|c|c|c|c|c|}
\hline
$k$ & Shell & $[r_{k-1},r_{r}]$ & $\alpha _{k}$ & $\beta _{k}$ & $\gamma
_{k} $ \\ \hline
1 & Inner core & [0, 0.192] & 6.099 & -4.119 & 0.00 \\
2 & Outer core & [0.192,0.546] & 5.803 & -3.653 & -1.086 \\
3 & Lower mantle & [0.546,0.895] & 3.156 & -1.459 & 0.280 \\
4 & Transition & [0.895,0.937] & -5.376 & 19.210 & -12.520 \\
5 & Upper mantle & [0.937,1] & 11.540 & -20.280 & 10.410 \\ \hline
\end{tabular}%
\end{center}
\caption{Coefficients of the electron density parametrization $N_{k}(r)=%
\protect\alpha _{k}+\protect\beta _{k}r^{2}+\protect\gamma
_{k}r^{4}$ in mol/cm$^{3}$, for the different shells. The radial
distance $r$ is normalized to the Earth radius. The fitted values
are taken from Ref. \protect\cite{2v}.}
\end{table}
\begin{figure}[tbp]
\includegraphics[width=8cm]{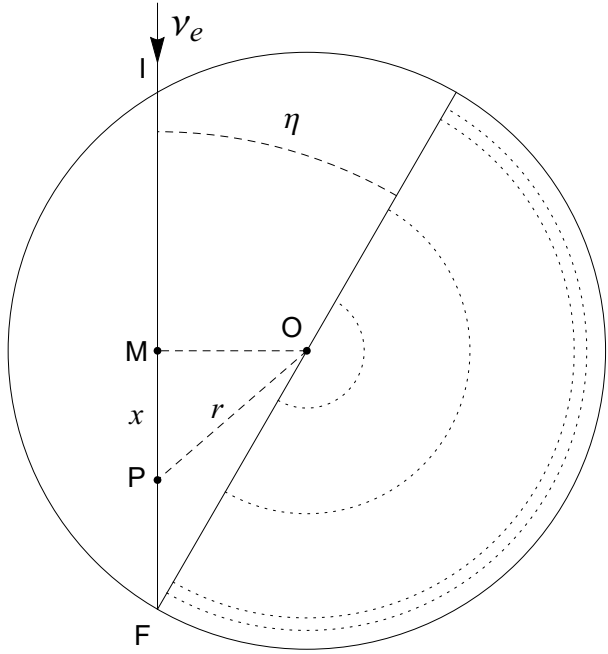}
\bigskip
\caption{Cross section of the Earth showing five shells. I and F are
entering and exit points of neutrino trajectory respectively, M is
the trajectory mid point, $r$ is the radial distance, and $x$ is
distance from mid point M.}
\end{figure}
\section{Application of perturbation theory}

\noindent In this section we present the analytic expression of neutrino
evolution operator $\mathcal{U}(x)$ for propagation of neutrino through the
Earth interior. Using equation (\ref{nx}), we split the Hamiltonian $\tilde{H%
}^{(f)}(x)$ given in equation (\ref{hft}) for the $k$th shell as
following
\begin{equation}
{\tilde{H}^{(f)}(x)=\bar{H}+\delta \tilde{H}^{(f)}(x),}
\end{equation}%
where $\bar{H}=\tilde{H}^{(f)}|_{N_{e}(x)\rightarrow \bar{N}_{k}}$ and $%
\delta \tilde{H}^{(f)}(x)$ is
\begin{equation}
{\delta \tilde{H}^{(f)}(x)=\frac{2}{3}\sqrt{2}G_{F}}\mathrm{Diag}{[2\delta N}%
_{k}{(x),-\delta N}_{k}{(x),-\delta N}_{k}{(x)].}
\end{equation}%
This splitting insure that the constant part of the Hamiltonian
$\bar{H}$ is traceless so that we could use the expressions of
constant density given in Sec. III. Since the variation of density
in each shell is relatively small so we can treat $\delta
\tilde{H}^{(f)}(x)$ part of Hamiltonian as a perturbation and use
the following perturbative solution of evolution
operator of Eq. (\ref{evo}) for $k$th shell%
\begin{equation}
{\mathcal{U}(x_{k}-x_{k-1})=e^{-i\bar{H}(x_{k}-x_{k-1})}-i%
\int_{x_{k-1}}^{x_{k}}dxe^{-i\bar{H}(x_{k}-x)}\delta \tilde{H}%
^{(f)}(x)e^{-i\bar{H}(x-x_{k-1})}.}  \label{40}
\end{equation}%
Using Eq. (\ref{fut1}), which gives evolution operator for constant density,
we can calculate the exponential factors containing constant density
Hamiltonian $\bar{H}_{k}$ in equation (\ref{40}). The resultant expression
is given by
\begin{equation}
{\mathcal{U}(x_{k}-x_{k-1})}{=\frac{1}{3}\sum_{\alpha =1}^{3}e^{-iE_{\alpha
}^{(m)}(x_{k}-x_{k-1})}A_{\alpha }}{-\frac{i}{9}}\sum_{\alpha ,\beta }{%
\int_{x_{k}-1}^{x_{k}}dxe^{-iE_{\alpha }^{(m)}(x_{k}-x)}A_{\alpha }\delta
\tilde{H}^{(f)}(x)A_{\beta }e^{-iE_{\beta }^{(m)}(x-x_{k-1})}.}  \label{41}
\end{equation}%
Using this equation we can calculate evolution operator of neutrino
evolution in each shell. Total evolution operator is obtained from
the product of evolution operators for all shells coming in neutrino
trajectory for any given $\eta$, as following
\begin{equation}
{\mathcal{U}(x_{F},x_{I})=\prod_{k}\mathcal{U}(x_{k}-x_{k-1}).}  \label{42}
\end{equation}%
The expression (\ref{41}) is our main result of evolution operator for three
neutrino oscillations through the Earth interior.

\section{Results and discussion}

\noindent Using the analytical expression of evolution operator through
Earth interior, given in Eq. (\ref{42}), we can calculate the probability $%
P(v_{\alpha }\rightarrow v_{\beta })=\left\vert {\mathcal{U}}_{\beta \alpha }%
{(x_{F},x_{I})}\right\vert ^{2}$ for given nadir angle and oscillations
parameters that include three mixing angles $(\theta _{12},\theta
_{23},\theta _{13})$, one Dirac CP\ phase $\delta _{CP}$, and two
independent squared mass differences $(\Delta m_{21}^{2},\Delta m_{32}^{2})$%
. In this section we compare the results of analytical expression with the
solutions obtained by numerically solving the evolution Eq. (\ref{evo})
through the Earth interior. For comparison, we use three neutrino
oscillation parameters of LMA solution (i.e., $\sin ^{2}\theta
_{12}=0.307,\sin ^{2}\theta _{23}=0.417,\sin ^{2}\theta _{13}=2.12\times
10^{-2},\Delta m_{21}^{2}=7.49\times 10^{-5}$ eV$^{2},\Delta
m_{32}^{2}=2.51\times 10^{-3}$ eV$^{2},\delta _{CP}\approx 1.45\pi $). For
solar neutrinos $v_{e}$ the effect of the Earth matter interaction can be
described by day-night asymmetry of average electron survival probability.
For 3 neutrinos the day-night difference of averaged electron neutrino
survival probability is given by%
\begin{equation}
P_{N}-P_{D}=-\cos ^{2}\theta _{13}\overline{\cos 2\widehat{\theta }_{12}}%
\left( \left\langle P_{2e}^{\oplus }\right\rangle -P_{2e}^{(0)}\right)
\label{dn}
\end{equation}%
\noindent where $\cos 2\widehat{\theta }_{12}$ is the effective value of $%
\cos 2\theta _{12}$ at the production point of solar neutrino and
bar stands for production point averaged value,
$P_{2e}^{(0)}=\left\vert U_{e2}\right\vert ^{2}$, and $\left\langle
P_{2e}^{\oplus }\right\rangle $ is the conversion probability of
$v_{2}\rightarrow v_{e}$ through Earth interior averaged over
nadir angle, as following%
\begin{equation}
\left\langle P_{2e}^{\oplus }\right\rangle =\int\limits_{0}^{\pi /2}d\eta
W(\eta )P_{2e}^{\oplus },
\end{equation}%
\noindent where $W(\eta )$ is nadir angle distribution function,
whose functional form depends on latitude of the detector. We use
the analytical expression of $W(\eta )$ given in Ref. \cite{2v}. The
Eq. (\ref{dn}) shows that the effect of the Earth matter interaction
is described by non-zero value of $P_{2e}^{\oplus }-P_{2e}^{(0)}$.
In Fig. 2, the plot of $P_{2e}^{\oplus }-P_{2e}^{(0)}$ versus $\cos
{\eta }$ is given at $E=10$ MeV. The solid curve in the figure
represents the solution obtained by numerically solving neutrino
evolution equation and marker symbols represent the results obtained
from analytical expression. Disk (red) and rectangular (green) marks
represent values with and without first order correction
respectively. It is noted that the first order correction include
the effect of variation of the Earth density in each shell, whereas
in the results without the correction Earth density in each shell in
treated constant. The comparison of the results given in Fig. 2
shows that constant density approximation fails for $\cos \eta
>0.5$, i.e., when neutrinos total path length in the Earth is
greater than its radius. The plot also shows that our analytical
expression with first order correction nicely agree with numerical
solution for all values of nadir angle. In Fig. 3, we plot nadir
angle averaged value of $P_{2e}^{\oplus }-P_{2e}^{(0)}$ versus
energy relevant for solar neutrinos. The detector location is
assumed to be at Kamioka. The plot shows excellent agreement of
analytical results (both with and without first order correction)
with numerical solutions. This implies that treating the Earth
density piecewise constant in five shells is a good approximation as
far as calculations of averaged probability is concerned. The plot
of Fig. 3 also shows that day-night asymmetry effect is very weak
for solar neutrinos. The effect is not expected to be measured in
SNO or BOREXINO experiments, as it is below their sensitivity
\cite{em2004}.

\begin{figure}[tbp]
\includegraphics[width=16cm]{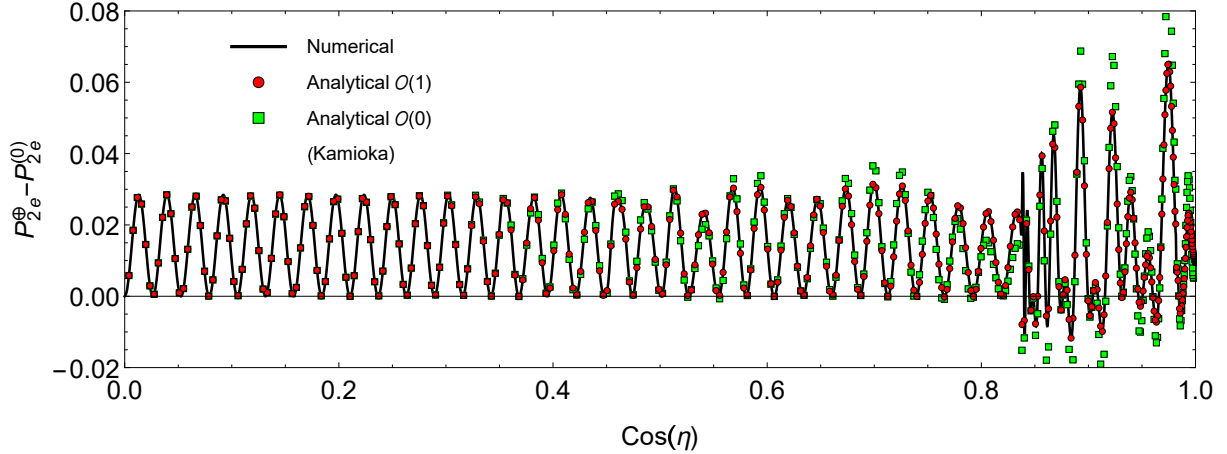}
\bigskip
\caption{Comparison of $P_{2e}^{\oplus }-P_{2e}^{(0)}$ versus $%
\cos \protect\eta $ at $E=10$ MeV. Solid curve represents the values
obtained from numerical solution of evolution Eq.
(\protect\ref{evo}), whereas disk (red) and rectangular (green)
marks represent the values obtained from our analytical expression
with and without first order correction respectively. Oscillation
parameters are taken of 3 neutrino LMA solution. }
\end{figure}

\begin{figure}[tbp]
\includegraphics[width=12cm]{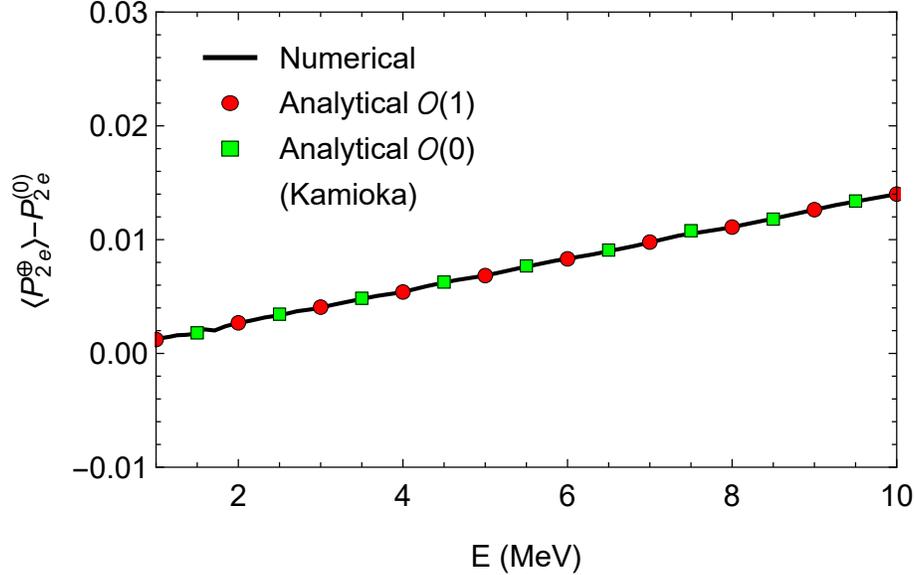}
\bigskip
\caption{Comparison of nadir angle averaged values of
$P_{2e}^{\oplus }-P_{2e}^{(0)}$ versus energy relevant for solar
neutrinos.}
\end{figure}

In Fig. 4, we plot the conversion probability $P(v_{\mu }\rightarrow v_{e})$
versus energy relevant for atmospheric and accelerator neutrinos. We produce
three plots corresponding to three values of $\eta =0^{\text{o}},40^{\text{o}%
},$ and $80^{\text{o}}$. In each plot a comparison of analytical
results (with and without first order correction) is made with the
results obtained from numerical solutions. The plots show excellent
agreement of first order corrected analytical results with the
numerical solutions. A substantial increase in the probability
observed about energy $E=5$ GeV is due to parametric enhancement
well studied in Refs. \cite{penha,penha1,penha2}. As
mentioned in Sec. II that the signs of electron density $N_{e}$ and CP phase $%
\delta _{CP}$ are inverted in the evolution equation when applied on
anti-neutrinos. As a result, the conversion probability
$P(\overline{v}_{\mu }\rightarrow \overline{v}_{e})$ becomes
different from $P(v_{\mu }\rightarrow v_{e})$. A part of this change
is due to inverting the sign of density and a part due to inverting
the sign of CP phase. In Fig. 5, we plot nadir angle averaged
conversion probabilities $\left\langle P(v_{\mu }\rightarrow
v_{e})\right\rangle $ and $\left\langle P(\overline{v}_{\mu
}\rightarrow \overline{v}_{e})\right\rangle $ versus energy using
our analytical expression. The plots clearly show a relative
suppression in the
probability $\left\langle P(\overline{v}_{\mu }\rightarrow \overline{v}%
_{e})\right\rangle $. In the same figure we also plot the conversion
probabilities assuming zero density. These plots (dashed line) show that a
change of about 0.05 in the value of conversion probabilities of $v_{\mu
}\rightarrow v_{e}$ and $\overline{v}_{\mu }\rightarrow \overline{v}_{e}$ is
produced due to inverting the sign of CP\ phase in the energy range $0.5$ to
$10$ GeV, if density is taken zero.

Recently T2K experiment \cite{T2Kcp} has measured CP phase through
the study of difference between the conversion probabilities of
$v_{\mu }\rightarrow v_{e}$ and $\overline{v}_{\mu }\rightarrow
\overline{v}_{e}$ using accelerator neutrinos. In the base-line
length of $295$ km the expected change in the $P(v_{\mu }\rightarrow
v_{e})$ and $P(\overline{v}_{\mu }\rightarrow \overline{v}_{e})$
probabilities is less than 0.025 for $E>0.4$ GeV. More importantly,
we find that the change is minimally sensitive to $\delta_{CP}$,
which makes precise measurement of $\delta _{CP}$ difficult. It is
noted that for base-line length 295 km, the Earth matter effect is
negligible. The sensitivity to CP phase is energy dependent and can
also be affected by the Earth matter effect. This effect is
highlighted in the Fig. 6, in which we plot the change in nadir
angle averaged probabilities $\left\langle P(v_{\mu }\rightarrow
v_{e})\right\rangle$ and $\left\langle P(\overline{v}_{\mu
}\rightarrow \overline{v}_{e})\right\rangle$ versus energy for
three different values of $\delta _{CP}$. The central curve correspond to $%
\delta _{CP}=1.45\pi $; the best fitted value obtained by T2K
experiment \cite{T2Kcp}, whereas upper and lower curves correspond
to the values 1.1$\pi $ and 1.8$\pi $, defined by measured error
limits of $\delta _{CP}$ in T2K experiment. The plots clearly show
that sensitivity to CP phase is indeed affected by Earth matter
effect and in energy range 0.2 to 1 GeV,
sensitivity to CP phase is maximum. In this energy range a variation of $%
\delta _{CP}$ in the range 1.1$\pi $ to 1.8$\pi $ can produce a
variation of 0.1
in $\left\langle P(v_{\mu }\rightarrow v_{e})\right\rangle -\left\langle P(%
\overline{v}_{\mu }\rightarrow \overline{v}_{e})\right\rangle $. In
order to study how the sensitivity to CP phase changes with nadir
angle, we plot energy averaged probability difference $\left\langle
P(v_{\mu }\rightarrow v_{e})\right\rangle _{E}-\left\langle
P(\overline{v}_{\mu }\rightarrow \overline{v}_{e})\right\rangle
_{E}$ versus nadir angle in Fig. 7. The energy averaged conversion
probabilities $\left\langle P(v_{\mu }\rightarrow
v_{e})\right\rangle _{E}$ and $\left\langle P(\overline{v}_{\mu
}\rightarrow \overline{v}_{e})\right\rangle _{E}$ are calculated
using Gaussian energy spectrum. The position of peak ${\normalsize
E=0.6}$ GeV {\normalsize and its width }$\sigma \approx 0.3$ GeV are
approximated from the energy spectrum of $v_{\mu }$ and
$\overline{v}_{\mu }$ in T2K accelerator neutrino experiment
\cite{T2Kcp}. In Fig. 7, we plot {\normalsize $\left\langle
P(v_{\mu }\rightarrow v_{e})\right\rangle _{E}-\left\langle P(\overline{v}%
_{\mu }\rightarrow \overline{v}_{e})\right\rangle _{E}$ versus nadir
angle corresponding to same three values of} {\normalsize $\delta
_{CP}$ used in Fig. 6. The plots show that sensitivity to $\delta
_{CP}$ varies with nadir angle and it is maximum about }$\cos \eta
=0.3$ ($\eta =72.5^{\text{o}}$), which corresponds to base-line
length $L=3827$ km. At the maxima, a variation of {\normalsize
$\delta _{CP}$ in the range 1.1$\pi $ to 1.8$\pi $} can produce a
variation of 0.07 in {\normalsize $\left\langle P(v_{\mu
}\rightarrow v_{e})\right\rangle _{E}-\left\langle
P(\overline{v}_{\mu }\rightarrow \overline{v}_{e})\right\rangle
_{E}$.} It is noted that this variation is merely 0.0022 at
$\cos\eta=0.023$ ($\eta =88.7^{\text{o}}$); the nadir angle of T2K
experiment. This suggests that an accelerator neutrino experiment
constructed to work at base-line length $L\simeq 3872$ km could be
ideal for precise measurement of {\normalsize $\delta _{CP}.$ }

\begin{figure}[tbp]
\includegraphics[width=8cm]{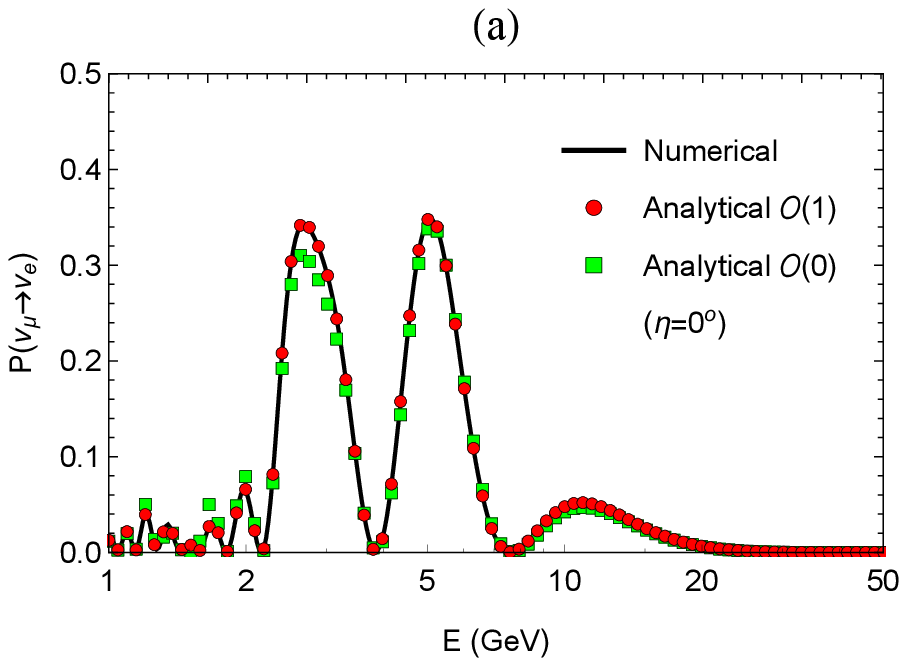} \includegraphics[width=8cm]{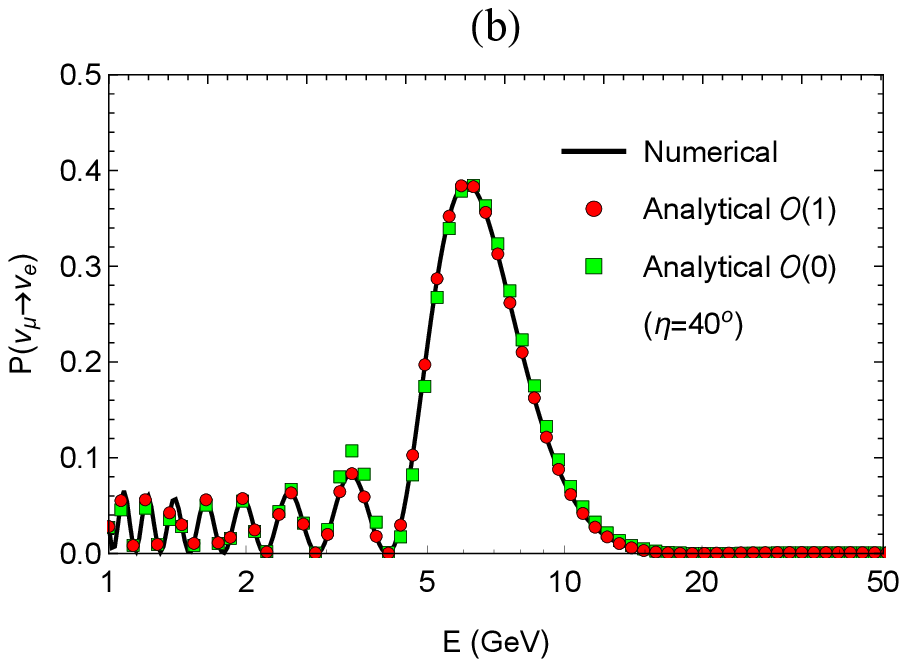}
\includegraphics[width=8cm]{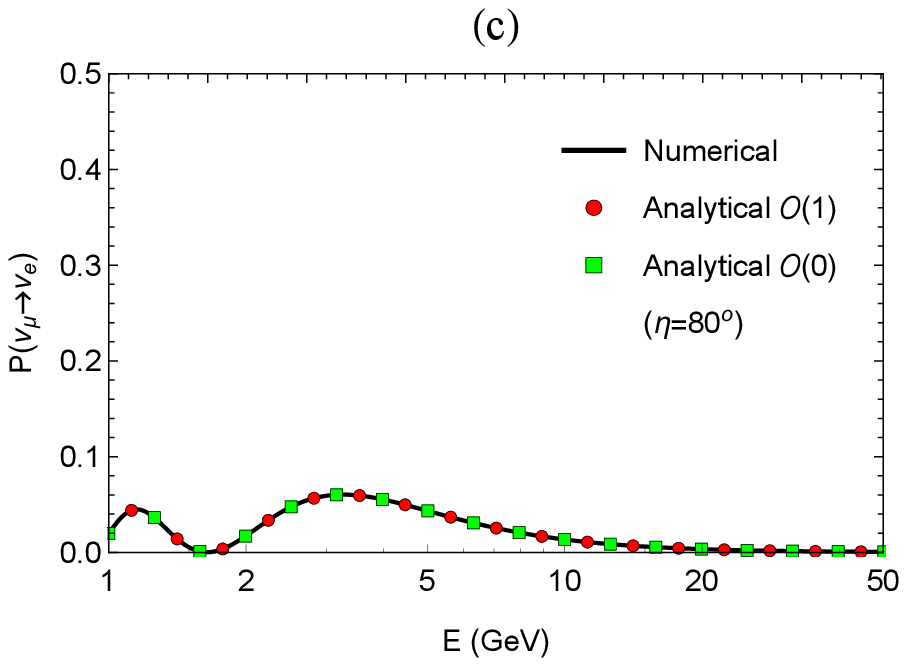}
\bigskip
\caption{Comparison of conversion probability $P(v_{\protect\mu
}\rightarrow v_{e})$ versus energy relevant for atmospheric and
accelerator neutrinos for
three different values of nadir angle a) $\protect\eta =0^{\text{o}},$ b) $%
\protect\eta =40^{\text{o}},$ and c) $\protect\eta =80^{\text{o}}.$}
\end{figure}

\begin{figure}[tbp]
\includegraphics[width=12cm]{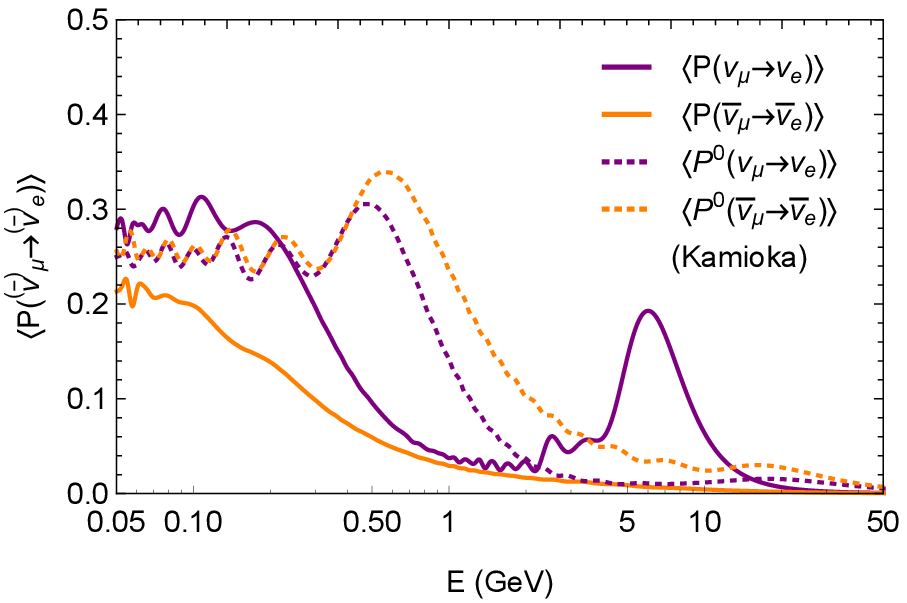}
\bigskip
\caption{Plots of nadir angle averaged survival probabilities
$\left\langle P(v_{\mu }\rightarrow v_{e})\right\rangle$ and
$\left\langle P(%
\overline{v}_{\mu }\rightarrow \overline{v}_{e})\right\rangle$
versus energy. Upper (purple) and lower (orange) solid curves
represent probabilities $P(v_{\protect\mu }\rightarrow v_{e})$ and $P(%
\overline{v}_{\protect\mu }\rightarrow \overline{v}_{e})$ respectively. The
dashed curves, produced for comparison, represent the same probabilities
assuming zero density. }
\end{figure}
\bigskip

\begin{figure}[tbp]
\includegraphics[width=12cm]{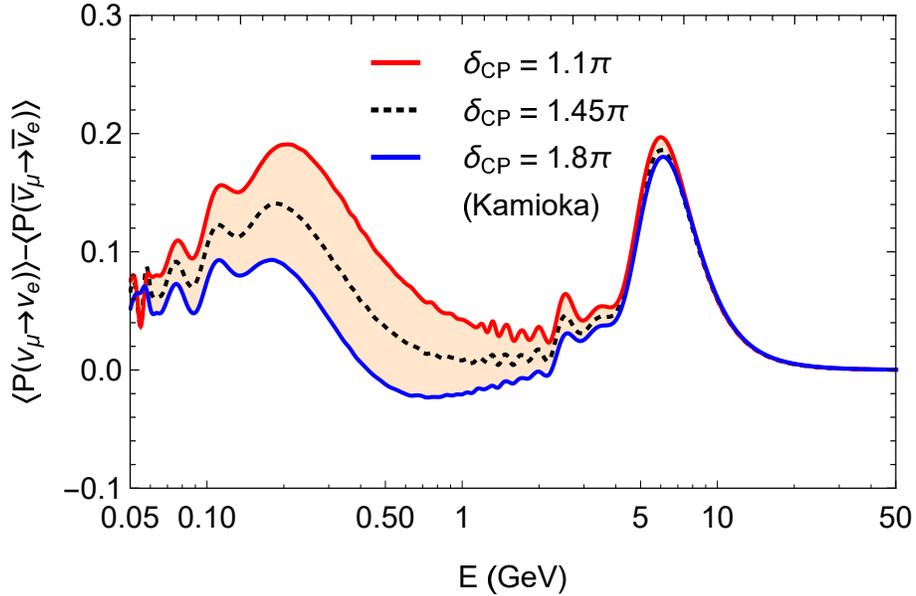}
\bigskip
\caption{Plots of nadir angle averaged probability difference
$\left\langle
P(v_{\protect\mu }\rightarrow v_{e})\right\rangle -\left\langle P(\overline{v%
}_{\protect\mu }\rightarrow \overline{v}_{e})\right\rangle $ versus energy
for three different values of $\protect\delta _{CP}$. For upper (red) and
lower (blue) solid curves $\protect\delta_{CP}$ is $1.1\protect\pi$ and $1.8%
\protect\pi$ respectively. }
\end{figure}

\begin{figure}[tbp]
\includegraphics[width=12cm]{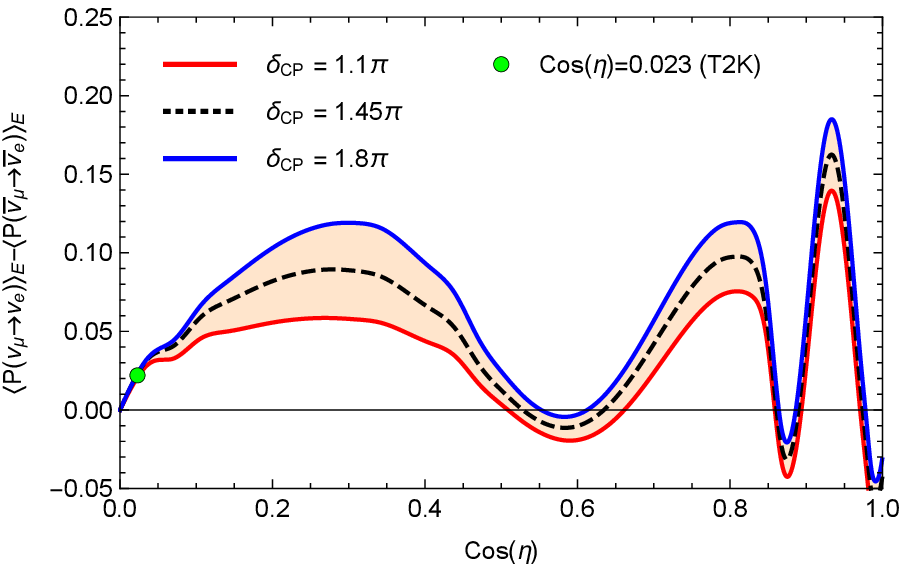}
\bigskip
\caption{Plots of energy averaged probability difference $\left\langle P(v_{%
\protect\mu }\rightarrow v_{e})\right\rangle _{E}-\left\langle P(\overline{v}%
_{\protect\mu }\rightarrow \overline{v}_{e})\right\rangle _{E}$
versus $\cos \protect\eta $ for three different values of
$\protect\delta _{CP}$. The circular mark (green) is at $\cos
\protect\eta =0.023$, corresponding to base-line length $L=295$ km
of T2K experiment.   }
\end{figure}

\acknowledgements{FA acknowledge the financial support of HEC
through research grant 20-4500/NRPU/R\&D/HEC/14/727 }

\end{document}